\documentclass[aps,prb,twocolumn,longbibliography,showpacs,superscriptaddress]{revtex4-1}

\usepackage[T1]{fontenc}
\usepackage[utf8]{inputenc}
\usepackage{graphicx}
\usepackage{amsmath}
\usepackage{dsfont}
\usepackage{amssymb}
\usepackage{bm}
\usepackage{siunitx}
\usepackage{physics}
\usepackage{xcolor}
\usepackage{hyperref}
\usepackage[normalem]{ulem}
\usepackage{placeins}

\AtBeginDocument{%
    \renewcommand{\abstractname}{Abstract}%
}
\makeatletter
\renewcommand{\frontmatter@abstractheading}{%
    \begingroup
    \centering
    \normalfont\footnotesize
    Corresponding author: Victor Pardo
    (\href{mailto:victor.pardo@usc.es}{victor.pardo@usc.es})\par
    \medskip
    \normalfont\normalsize\bfseries \abstractname\par
    \medskip
    \endgroup
}
\makeatother

\hypersetup{
    colorlinks=true,
    linkcolor=blue,
    citecolor=blue,
    filecolor=blue,
    urlcolor=blue
}

\begin{document}

\title{Magnetoelectric flat band induced by a $\sqrt{3}\times\sqrt{3}$ charge density wave in monolayer CrSe$_2$}

\author{Pablo Savino}
\affiliation{Departamento de Física Aplicada, Universidade de Santiago de Compostela, E-15782 Santiago de Compostela, Spain}
\affiliation{Instituto de Materiais iMATUS,
Universidade de Santiago de Compostela, E-15782 Campus Sur s/n,
Santiago de Compostela, Spain}

\author{M. del Carmen Fuente Santiago}
\affiliation{Departamento de Física Aplicada, Universidade de Santiago de Compostela, E-15782 Santiago de Compostela, Spain}
\affiliation{Instituto de Materiais iMATUS,
Universidade de Santiago de Compostela, E-15782 Campus Sur s/n,
Santiago de Compostela, Spain}

\author{Jan Phillips}
\affiliation{International Iberian Nanotechnology Laboratory (INL),
Av. Mestre José Veiga, 4715-330 Braga, Portugal}

\author{Javier Corral Sertal}
\affiliation{Departamento de Física Aplicada, Universidade de Santiago de Compostela, E-15782 Santiago de Compostela, Spain}
\affiliation{Center for Research in Biological Chemistry and Molecular Materials CIQUS,
Universidade de Santiago de Compostela, E-15782 Campus Sur s/n,
Santiago de Compostela, Spain}

\author{Adolfo O. Fumega}
\affiliation{Department of Applied Physics, Aalto University, 02150 Espoo, Finland}

\author{Santiago Blanco-Canosa}
\affiliation{Donostia International Physics Center (DIPC),
Paseo Manuel de Lardizábal, 20018 San Sebastián, Spain}
\affiliation{IKERBASQUE, Basque Foundation for Science,
48013 Bilbao, Spain}

\author{Victor Pardo}
\affiliation{Departamento de Física Aplicada, Universidade de Santiago de Compostela, E-15782 Santiago de Compostela, Spain}
\affiliation{Instituto de Materiais iMATUS,
Universidade de Santiago de Compostela, E-15782 Campus Sur s/n,
Santiago de Compostela, Spain}

\begin{abstract}
We investigate from first principles the metastable polar
$\sqrt{3}\times\sqrt{3}$ trimerized phase of monolayer CrSe$_2$,
which hosts an isolated, nearly dispersionless band at the Fermi level
in the majority-spin channel. Combining orbital-resolved \textit{ab initio}
calculations with an effective tight-binding model, we show that the flat
band originates from the molecular electronic reconstruction induced by
Cr trimerization rather than from on-site Coulomb interactions. The trimerized reconstruction is accompanied by an asymmetric
out-of-plane displacement of the Se sublattice, which breaks inversion
symmetry and generates an intrinsic polarity perpendicular to the
monolayer. Exploiting this polar character, we show that a perpendicular
electric field can promote access to the
$\sqrt{3}\times\sqrt{3}$ phase and, once formed, further tune the
dispersion of its narrow electronic band. Our results reveal a
structurally driven and electrically controllable route to flat-band
engineering in a magnetic two-dimensional material.
\end{abstract}

\maketitle
\raggedbottom

\begin{figure*}[t]
    \centering
    \includegraphics[width=\textwidth]{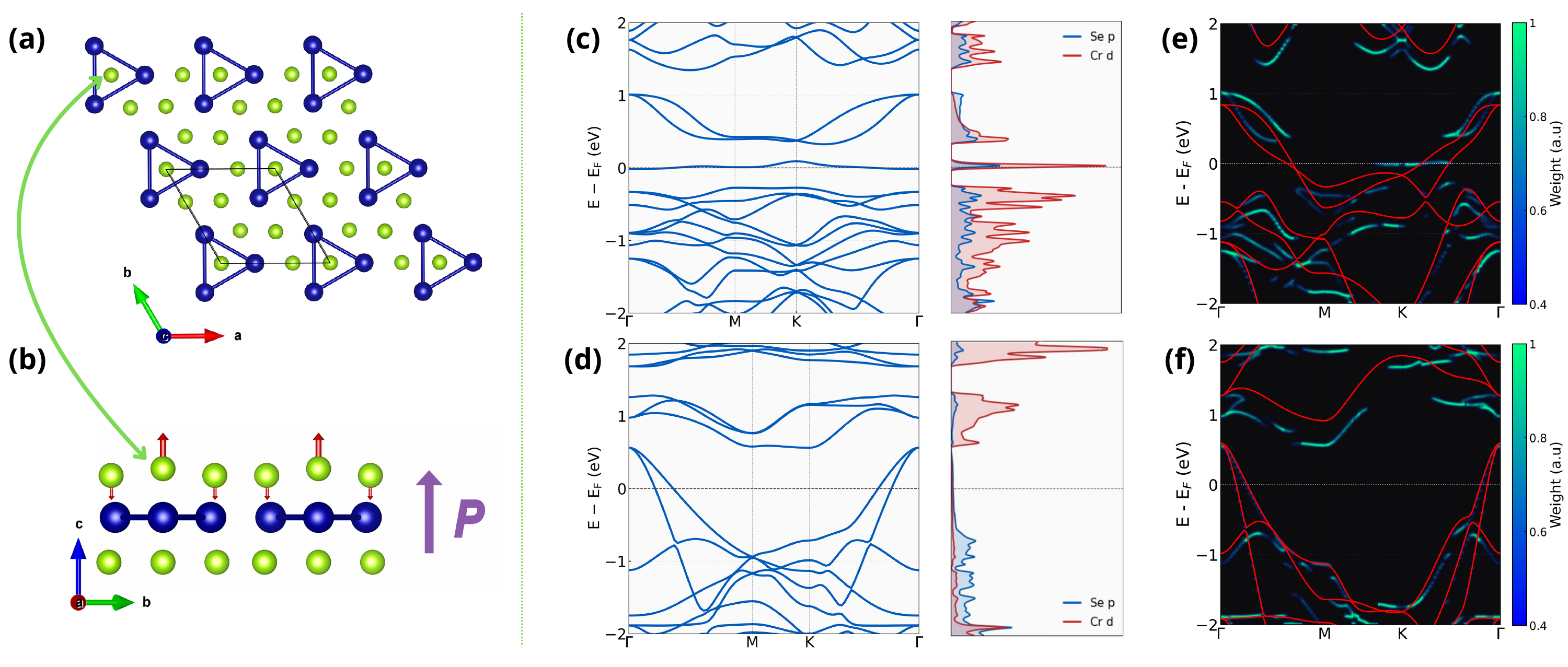}
    \caption{
    Structural and electronic properties of the polar
    $\sqrt{3}\times\sqrt{3}$ trimerized phase of monolayer CrSe$_2$.
    (a) Top view showing the formation of Cr trimers.
    (b) Side view highlighting the out-of-plane displacement of the Se atoms
    and the resulting polarization.
    (c,d) Spin-resolved band structures and projected density of states for
    the majority- and minority-spin channels, respectively, showing the
    isolated nearly flat band at the Fermi level in the majority-spin channel.
    (e,f) Corresponding unfolded band structures projected onto the
    primitive 1T Brillouin zone for the majority- and minority-spin channels,
    respectively. The red lines show the band structure of the undistorted
    1T phase.
    }
    \label{fig:electronic_structure}
\end{figure*}
\begin{figure*}[t]
    \centering
    \includegraphics[width=\textwidth]{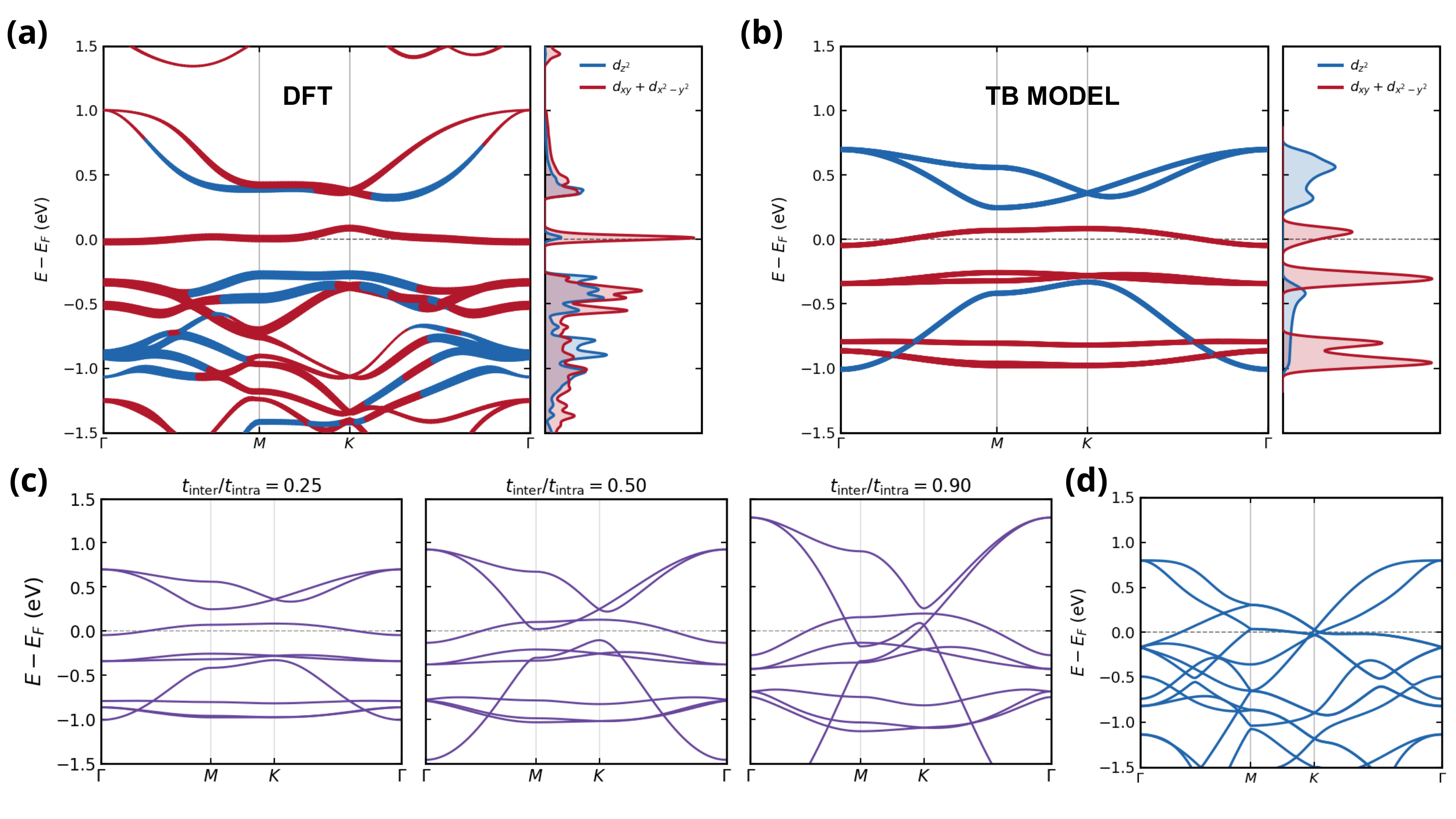}
    \caption{
Electronic structure and tight-binding description of the
$\sqrt{3}\times\sqrt{3}$ trimerized phase of monolayer CrSe$_2$.
(a) Majority-spin DFT band structure and projected density of states of the
trimerized phase, resolved into the $d_{z^2}$ contribution and the combined
in-plane $d_{xy}+d_{x^2-y^2}$ contribution. The isolated nearly flat band at
the Fermi level is dominated by the in-plane orbitals.
(b) Band structure and orbital-resolved density of states obtained from the
effective three-orbital tight-binding model, with $t_{\mathrm{inter}}/t_{\mathrm{intra}}=0.25$, showing the same isolated
in-plane-derived nearly flat band.
(c) Evolution of the tight-binding band structure upon progressively
increasing the inter-trimer hopping relative to the intra-trimer hopping,
$t_{\mathrm{inter}}/t_{\mathrm{intra}}=0.25$, $0.50$, and $0.90$.
The bands become increasingly dispersive as the distinction between intra-
and inter-trimer couplings is reduced.
(d) Majority-spin DFT band structure of the undistorted 1T phase, where the
isolated flat-band structure characteristic of the
$\sqrt{3}\times\sqrt{3}$ trimerized phase is absent.
}
    \label{fig:flat_band_model}
\end{figure*}

\section*{Introduction}

Two-dimensional (2D) materials offer a fertile ground for quantum phenomena often absent in bulk systems owing to reduced dimensionality, strong electronic correlations, and tunable lattice geometries.\cite{novoselov20162d} Of particular interest are systems hosting nearly flat electronic bands near the Fermi level, where kinetic energy is strongly suppressed, amplifying Coulomb interactions and enabling unconventional correlated states---from superconductivity\cite{aoki2020theoretical,iglovikov2014superconducting,chan2022designer} and magnetism\cite{tasaki1998nagaoka} to enhanced correlation effects.\cite{checkelsky2024flat}

Flat bands arise in diverse platforms: kagome lattices leverage geometric frustration to host charge order,\cite{neupert2022charge,teng2022discovery} spin-density waves,\cite{chen2024intertwined} and superconductivity;\cite{wilson2024v3sb5} twisted van der Waals heterostructures (e.g., twisted bilayer graphene) generate moiré-induced flat bands that support Mott-like insulators\cite{cao2018correlated} and unconventional superconductivity;\cite{cao2018unconventional} and certain transition metal dichalcogenides (TMDs)---even without moiré patterns---exhibit flat or weakly dispersive bands via orbital hybridization,\cite{phillips2024self} electron--phonon coupling,\cite{bianco2020weak} or structural distortions.\cite{neto2001charge} Notably, 1T-TaS$_2$,\cite{vavno2021artificial,crippa2024heavy} 1T-TaSe$_2$,\cite{dalal2025flat,wan2023evidence} and 1T-TiSe$_2$\cite{sugawara2016unconventional} display CDW order, Mott physics, and pressure- or doping-induced superconductivity,\cite{sipos2008mott,liu2013superconductivity} linked to flat-band formation, in the case of the Ta compounds as a consequence of the particular CDW reconstruction (the so-called star-of-David pattern: a $\sqrt{13}\times\sqrt{13}$ supercell).

This has spurred the concept of ``flat-band engineering''---designing materials with dispersionless states near the Fermi level through symmetry control,\cite{ramachandran2017chiral,bae2023isolated} chemical tuning,\cite{hase2018computational} interlayer twisting,\cite{zhang2020flat} or external fields like strain\cite{wan2023topological} and gating.\cite{park2020gate} A key challenge lies in understanding how flat bands couple to spin, lattice, and orbital degrees of freedom to stabilize novel quantum phases.

Charge-density waves (CDWs)---periodic electronic and lattice modulations\cite{gruner2018density}---are central to this interplay. Driven by Fermi-surface nesting, electron--phonon coupling, or correlations,\cite{johannes2008fermi} CDWs in 2D materials frequently coexist or compete with superconductivity,\cite{balseiro1979superconductivity,lian2018unveiling} magnetism,\cite{fumega2019absence} and excitonic condensation.\cite{lian2019charge} In TMDs, CDWs can reconstruct the band structure, open pseudogaps, and even generate flat subbands or polar phases by breaking inversion symmetry.\cite{rossnagel2011origin,yuan2019room}

Chromium-based TMDs have recently emerged as compelling platforms
where magnetism, CDWs, and flat-band physics intersect
\cite{phillips2024ab}. CrSe$_2$ and CrTe$_2$ adopt layered structures
with polymorphism sensitive to synthesis and strain
\cite{li2021van,cui2024stoichiometry,liu2022structural}, while their
partially filled Cr $3d$ orbitals introduce intrinsic magnetism
\cite{freitas2015ferromagnetism,zhang2021room}. Strong $d$--$p$
hybridization further drives structural instabilities, enabling
competing magnetic, metallic, semiconducting, and CDW states
\cite{lv2015strain,freitas2013antiferromagnetism,otero2020controlled}.
In bulk 1T-CrSe$_2$, synchrotron X-ray diffraction revealed successive
structural transitions involving the formation of linear Cr trimers,
attributed to Peierls-like instabilities and interlayer Se--Se
interactions \cite{kobayashi2019linear}. In contrast, $\beta$-NMR
measurements found no pronounced anomaly associated with these
structural transitions, leaving the microscopic nature of the
low-temperature reconstruction still open \cite{Ticknor2020}.

Theoretical studies suggest monolayer CrTe$_2$ may host a $\sqrt{3}\times\sqrt{3}$ ``$d$1T'' distortion, where Cr atoms form trimers within the 1T lattice.\cite{otero2020controlled} This symmetry-lowering reconstruction strongly reorganizes the Cr $d$ manifold through intra-trimer hybridization and can generate weakly dispersive electronic states together with a polar metallic structure. Analogous distortions in MoTe$_2$ produce ferroelectricity via asymmetric metal--chalcogen displacements.\cite{yuan2019room,jindal2023coupled,hou2019strain} Realizing such a phase in a magnetic TMD like CrSe$_2$ would offer a unique 2D platform for coupled spin--lattice--charge phenomena.

Here, we use first-principles calculations to investigate a metastable
$\sqrt{3}\times\sqrt{3}$ trimerized phase of monolayer CrSe$_2$.
The reconstruction produces short and long Cr--Cr bonds, distorts the
Se sublattice, and breaks inversion symmetry, giving rise to an
intrinsic out-of-plane polarization. Remarkably, this polar phase hosts an
isolated, nearly dispersionless band pinned at the Fermi level in the
majority-spin channel. We combine orbital-resolved \textit{ab initio}
calculations with an effective tight-binding model to uncover the
microscopic origin of this flat band, and further analyze the effects
of electronic correlations and magnetic order. Finally, we show that
the polar character of the same structural reconstruction provides a
route to access and tune the trimerized flat-band state using an
out-of-plane electric field.

\enlargethispage{\baselineskip}
\section*{Results and Discussion}

\subsection*{Structural distortion and energetics of the
$\sqrt{3}\times\sqrt{3}$ phase}

We begin by briefly describing the structural properties and energetic
stability of the $\sqrt{3}\times\sqrt{3}$ phase of monolayer
1T-CrSe$_2$. In the undistorted structure, the Cr atoms form a
triangular lattice coordinated by edge-sharing CrSe$_6$ octahedra.
The $\sqrt{3}\times\sqrt{3}$ reconstruction reorganizes this lattice
into Cr trimers, with the Cr atoms moving toward one another and
forming short intra-trimer and longer inter-trimer Cr--Cr bonds, as
shown in Fig.~\ref{fig:electronic_structure}(a).

The Se sublattice undergoes a corresponding reconstruction. In
particular, the Se atom located at the center of each Cr trimer is
displaced away from the Cr plane, while the surrounding Se atoms move
in the opposite direction [Fig.~\ref{fig:electronic_structure}(b)].
This asymmetric out-of-plane displacement breaks inversion symmetry
and gives rise to an intrinsic polar axis perpendicular to the
monolayer.

Phonon calculations for the high-symmetry 1T phase reveal several
unstable modes at the Brillouin-zone boundary, indicating an intrinsic
tendency toward structural reconstruction.\cite{phillips2024ab}
Among the competing distortions, the $2\times2$ and
$\sqrt{3}\times\sqrt{3}$ phases are the most favorable. The $2\times2$
structure lies approximately $40$ meV/Cr below the undistorted 1T
phase, whereas the $\sqrt{3}\times\sqrt{3}$ phase is only about
$8$ meV/Cr higher in energy than the $2\times2$ reconstruction.
This small energy difference places the polar trimerized phase among
the low-energy competing instabilities of monolayer CrSe$_2$ and
motivates the detailed analysis of its electronic properties.

\subsection*{Electronic structure of the flat-band phase}

The electronic structure of the polar
$\sqrt{3}\times\sqrt{3}$ phase is shown in
Fig.~\ref{fig:electronic_structure}(c,d), together with the corresponding
projected density of states. The most striking feature is the appearance
of an isolated, nearly dispersionless band pinned at the Fermi level in
the majority-spin channel. This state produces a pronounced peak in the
density of states at $E_{\mathrm{F}}$ and has predominantly Cr $d$
character. In contrast, the minority-spin channel remains metallic,
with several dispersive bands crossing the Fermi level and a substantial
contribution from Se $p$ states.

The projected density of states also shows strong Cr $d$--Se $p$
hybridization throughout the low-energy manifold, indicating that a
simple ionic Cr$^{4+}$ ($d^2$) picture is not sufficient to describe
the electronic structure. In such an ionic limit, the Se $p$ states
would be fully occupied and only two electrons per Cr would populate the Cr
$d$ manifold. Instead, Se $p$-derived hole states are present around
the Fermi level, particularly in the minority-spin channel. This internal charge transfer
effectively self-dopes the Cr-derived states and contributes to placing
the narrow majority-spin band directly at the Fermi level.

To directly visualize the reconstruction induced by the
$\sqrt{3}\times\sqrt{3}$ modulation, we unfold the supercell bands onto
the primitive 1T Brillouin zone, as shown in
Fig.~\ref{fig:electronic_structure}(e,f). The red lines correspond to
the undistorted 1T phase and provide a direct reference for identifying
the changes induced by the trimerization. In the majority-spin channel,
several dispersive bands cross the Fermi level in the parent structure,
whereas the $\sqrt{3}\times\sqrt{3}$ reconstruction opens gaps at a
number of these crossings. This is particularly evident around the
$K$ point, although additional gaps are also visible along
$\Gamma$--$K$ and $\Gamma$--$M$. These features are consistent with
the coupling and hybridization of electronic states folded into the
reduced Brillouin zone by the $\sqrt{3}\times\sqrt{3}$ superlattice,
illustrating how the structural reconstruction modifies the low-energy
electronic structure of the parent 1T phase.

A further characteristic feature of the unfolded spectrum is the strong
concentration of spectral weight of the isolated flat band around the
$K$ region. While the existence of the flat band is already evident in
the folded band structure, the unfolding makes explicit its connection
to the electronic reconstruction produced by the
$\sqrt{3}\times\sqrt{3}$ periodicity.

\subsection*{Trimerization-driven origin of the flat band}\label{sec:flat_band_origin}

To understand the origin of the nearly dispersionless band in the
$\sqrt{3}\times\sqrt{3}$ phase, we first analyze its orbital character.
In the energy window shown in Fig.~\ref{fig:flat_band_model}(a), the
dominant Cr-$d$ contributions arise from the $d_{z^2}$ orbital and from
the in-plane $d_{xy}$ and $d_{x^2-y^2}$ orbitals. The latter are shown
together as the combined $d_{xy}+d_{x^2-y^2}$ contribution.
Remarkably, the isolated band crossing the Fermi level is almost
entirely dominated by these in-plane orbitals, suggesting that its weak
dispersion is closely connected to the reorganization of the in-plane
Cr-$d$ states induced by the trimerization.

To rationalize this behavior, we construct a minimal effective
tight-binding model containing the $d_{z^2}$, $d_{xy}$, and
$d_{x^2-y^2}$ orbitals on each of the three Cr atoms of the
$\sqrt{3}\times\sqrt{3}$ unit cell. Since the model is intended to
describe the majority-spin sector of the ferromagnetic DFT calculation,
spin is not included explicitly. We retain three hopping channels: a
negative $d_{z^2}$--$d_{z^2}$ hopping, $t_{z^2}<0$, and two
bond-dependent hoppings within the in-plane orbital subspace, with
$t_{\sigma}<0$ and $t_{\pi}>0$. Hybridization between $d_{z^2}$ and
the in-plane orbitals is neglected in this minimal description in order
to isolate the essential orbital mechanism. As shown in Fig.~\ref{fig:three_orbital} in Appendix~A (Supplementary Information), introducing such hybridization redistributes
the orbital weight among the bands but does not modify the existence or
the predominantly in-plane character of the isolated narrow band.

The trimerization is introduced by distinguishing intra- and
inter-trimer hopping amplitudes, with the latter taken to be a fraction
of the corresponding intra-trimer amplitudes.

The molecular-level structure of the model, where we use $t_{\mathrm{inter}}/t_{\mathrm{intra}}=0.25$,  provides a simple picture
of the resulting band organization depicted in
Fig.~\ref{fig:flat_band_model}(b). In the $d_{z^2}$ sector, in blue, one orbital
on each of the three Cr sites gives three molecular states. For three
equivalent orbitals on a triangular trimer with a negative
nearest-neighbor hopping, the expected splitting consists of a lower
bonding singlet and an upper doubly degenerate pair, giving the
characteristic $1$ (singlet) $+\,2$ (doublet) structure.

The situation is richer for the in-plane sector, where two orbitals per
Cr give six states. Because the $d_{xy}$ and $d_{x^2-y^2}$ orbitals are
directional, their hopping depends on the orientation of the Cr--Cr
bond through the $\sigma$ and $\pi$ channels. The threefold symmetry of
the trimer reorganizes these six states into two singlets and two
doublets. For the signs and relative magnitudes of $t_{\sigma}$ and
$t_{\pi}$ used here, they appear in the sequence $2+1+2+1$ with
increasing energy, with the lowest in-plane level forming a doubly
degenerate pair as shown in the in-plane bands of Fig.~\ref{fig:flat_band_model}(b). Most importantly, one of the in-plane singlets becomes
energetically separated from the remaining trimer states and lies close
to the Fermi level.

When the trimers are coupled, this molecular state forms a narrow band.
Because the inter-trimer hopping is substantially smaller than the
intra-trimer hopping, its dispersion remains weak, producing the
isolated nearly flat band shown in Fig.~\ref{fig:flat_band_model}(b).
The model therefore naturally accounts for the dominant
$d_{xy}+d_{x^2-y^2}$ character found in the DFT flat band. In fact, the
isolated flat band is already obtained within the reduced in-plane
$d_{xy}$--$d_{x^2-y^2}$ subspace, showing that the essential mechanism
is fully captured by the molecular reorganization of these orbitals
within the Cr trimers. The $d_{z^2}$ states are still needed to
reproduce the broader orbital structure of the DFT bands in the same
energy range, but they are not required to describe the formation of
the flat band itself.

The role of the trimerization can be followed directly by progressively
increasing the inter-trimer hopping relative to the intra-trimer
hopping. Figure~\ref{fig:flat_band_model}(c) shows the evolution of the
tight-binding spectrum for
$t_{\mathrm{inter}}/t_{\mathrm{intra}}=0.25$, $0.50$, and $0.90$.
As the two hopping scales become comparable, the distinction between
individual trimers is progressively reduced and the trimer-derived
states acquire a substantially larger dispersion. In particular, the
isolated nearly flat band characteristic of the strongly trimerized
regime is gradually lost as the system approaches the uniform-hopping
limit.

This evolution is consistent with the first-principles band structure
of the undistorted 1T phase shown in
Fig.~\ref{fig:flat_band_model}(d). In the absence of the
$\sqrt{3}\times\sqrt{3}$ trimerization, the corresponding bands are
considerably more dispersive and the isolated flat band at the Fermi
level is no longer present. The combined DFT and tight-binding results
therefore identify the formation of weakly coupled Cr-trimer molecular
states, predominantly derived from the in-plane $d_{xy}$ and
$d_{x^2-y^2}$ orbitals, as the origin of the isolated nearly flat band
in the $\sqrt{3}\times\sqrt{3}$ phase.

\subsection*{Effect of electronic correlations and magnetism}

\begin{figure*}[t]
    \centering
    \includegraphics[width=\textwidth]{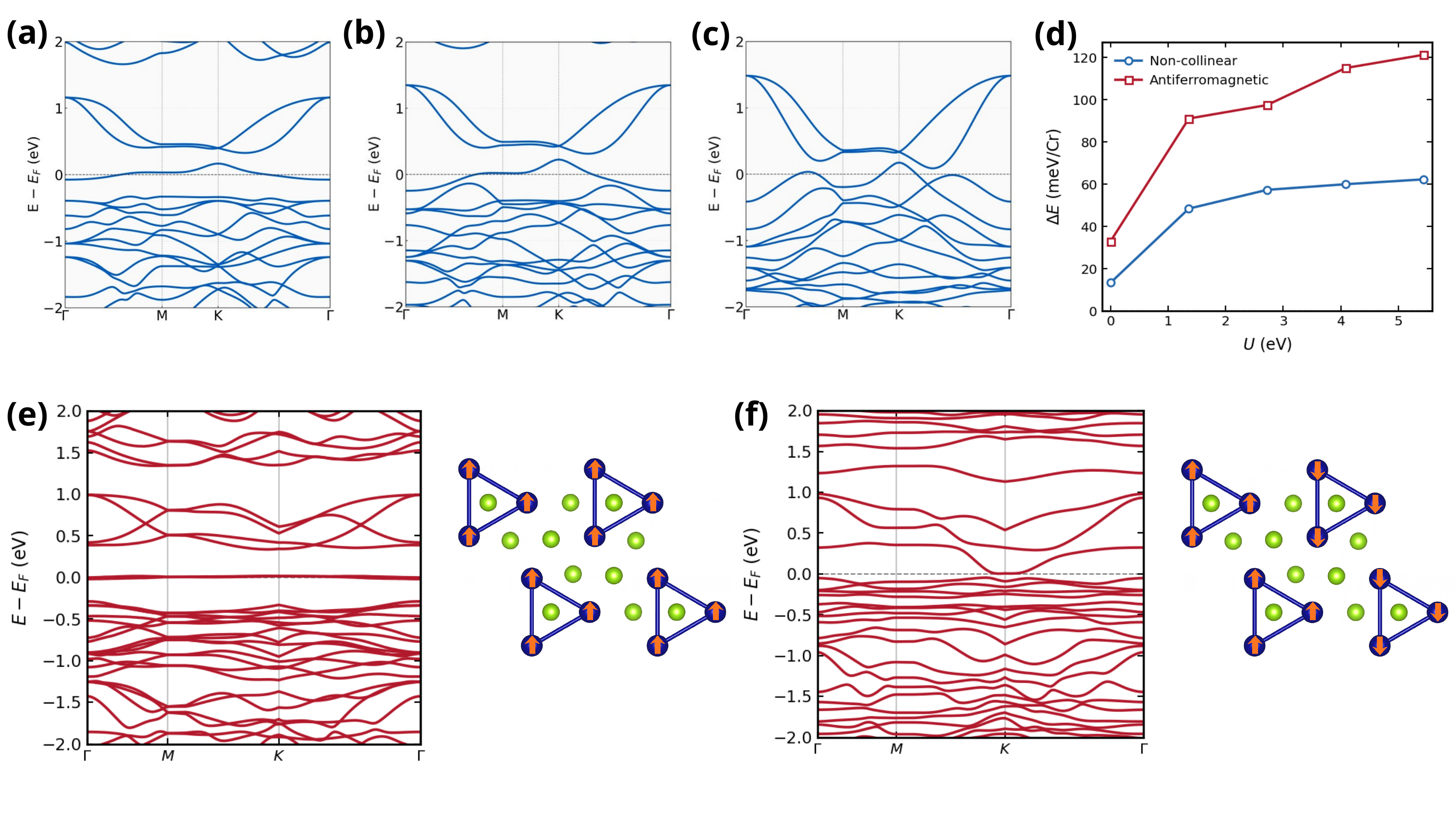}
    \caption{
    Effect of electronic correlations and magnetic order on the
    $\sqrt{3}\times\sqrt{3}$ trimerized phase of monolayer CrSe$_2$.
    (a--c) Majority-spin DFT+$U$ band structures for
    $U=1$, $2$, and $4$ eV, respectively, showing the evolution of the
    trimer-derived bands as the on-site Coulomb interaction is increased.
    The isolated nearly flat band is progressively modified and becomes
    more dispersive at larger $U$.
    (d) Energy differences between competing magnetic configurations as
    a function of $U$, taking the ferromagnetic state as the zero-energy
    reference. The non-collinear $120^\circ$ and collinear
    up--up--down (labeled as Antiferromagnetic) configurations remain higher in energy, with the
    ferromagnetic state becoming increasingly stabilized as $U$ increases.
    (e,f) Band structures for ferromagnetic and antiferromagnetic
    inter-trimer alignment, respectively, while preserving ferromagnetic
    order within each Cr trimer. The antiferromagnetic inter-trimer
    configuration lies approximately $58$ meV/Cr above the
    ferromagnetic one at $U=0$. The favored ferromagnetic configuration (e) retains
    the isolated nearly flat band at the Fermi level, whereas this
    characteristic feature is absent for antiferromagnetically (f) coupled
    trimers.
    }
    \label{fig:correlations_magnetism}
\end{figure*}
\begin{figure*}[t]
    \centering
    \includegraphics[width=\textwidth]{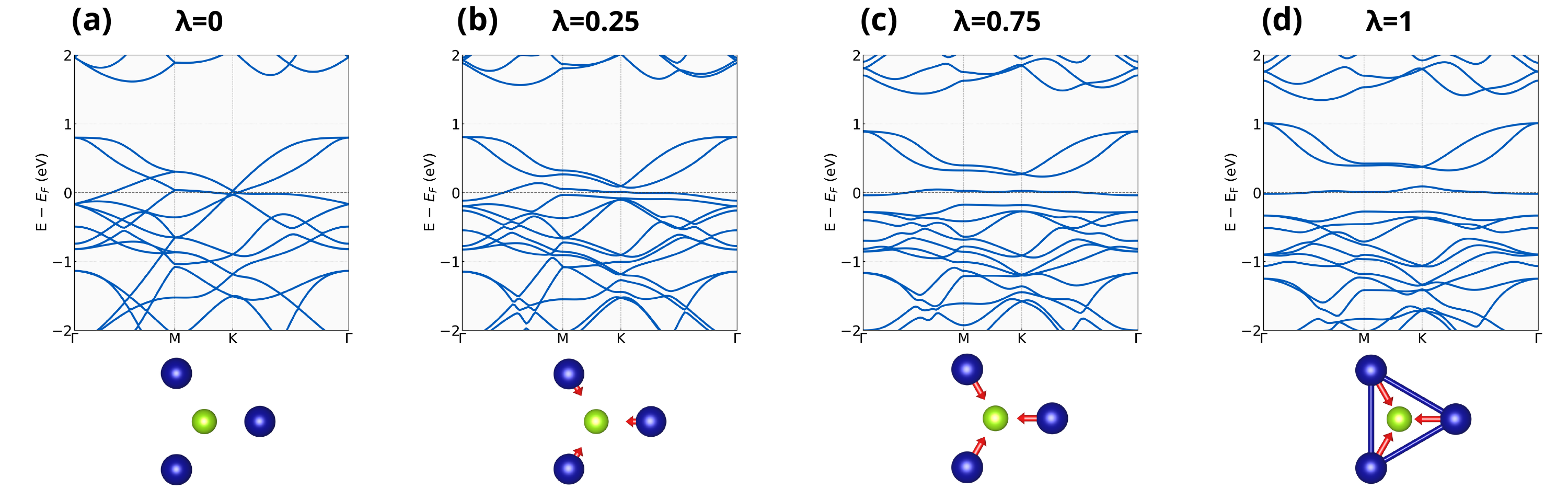}
    \caption{Evolution of the electronic structure along a continuous structural interpolation between the high-symmetry 1T phase and the fully developed polar $\sqrt{3}\times\sqrt{3}$ CDW phase of monolayer CrSe$_2$. (a--d) Majority-spin band structures for interpolation parameters $\lambda=0$, 0.25, 0.75, and 1, respectively. The corresponding structural sketches illustrate the progressive development of the CDW distortion, including Cr trimerization and the accompanying out-of-plane polar displacement of the Se sublattice. As the distortion develops, the low-energy Cr-$d$ manifold is progressively reconstructed and the isolated nearly flat band emerges near the Fermi level in the fully trimerized phase.}
   
    \label{fig:structural_path}
\end{figure*}

We next investigate how electronic correlations and magnetic order
affect the flat-band state of the $\sqrt{3}\times\sqrt{3}$ trimerized
phase. Figure~\ref{fig:correlations_magnetism}(a--c) shows the evolution
of the majority-spin band structure within DFT+$U$ for
$U=1$, $2$, and $4$ eV, respectively. At small and moderate values of
$U$, the characteristic narrow band at the Fermi level is still clearly
visible. As $U$ is increased, however, the Cr-$d$ bands undergo a
progressive reorganization and the initially narrow state becomes
increasingly dispersive.

This evolution can be understood from the molecular-orbital picture
introduced above. The formation of the flat band relies on the
intersite covalency and hybridization associated with the Cr trimers.
Increasing the on-site Coulomb interaction tends to localize the Cr-$d$
states and competes with this hybridization, progressively weakening
the electronic reconstruction associated with the trimerization. As a
result, the characteristic separation between the trimer-derived bands
becomes less pronounced and the electronic structure gradually
approaches the more dispersive band structure of the undistorted 1T
phase. Importantly, the flat band is already present in the absence of
a Hubbard correction and becomes less well defined as $U$ is increased.
This demonstrates that the flat band is not generated by on-site
Coulomb repulsion, but originates instead from the covalent
hybridization and molecular-state formation induced by the
trimerization. Electronic correlations can strongly modify this state,
but they are not its microscopic origin.

This raises the natural question of the effective correlation strength
in monolayer CrSe$_2$. Its metallic character and the strong Cr--Se
hybridization suggest that the material may lie in a moderately
correlated, hybridization-dominated regime, where the trimer molecular
states and the associated flat band remain well defined. Nevertheless,
a more strongly correlated regime cannot be excluded on the basis of
the present calculations alone. Experimental probes such as
angle-resolved photoemission spectroscopy (ARPES) or optical
spectroscopy would therefore be valuable for determining the degree of
electronic itinerancy and constraining the effective interaction
strength in monolayer CrSe$_2$.

The effect of electronic correlations on the magnetic stability is
shown in Fig.~\ref{fig:correlations_magnetism}(d). We compare the
ferromagnetic solution with two competing arrangements within the Cr
trimers: a collinear up--up--down configuration and a non-collinear
$120^\circ$ configuration. The ferromagnetic state remains the
lowest-energy solution throughout the entire range of $U$ considered.
Moreover, the energy separation with respect to both competing magnetic
configurations increases with $U$, indicating that electronic
correlations further stabilize ferromagnetic alignment in the
trimerized phase.

We finally examine the magnetic coupling between different Cr trimers.
Figures~\ref{fig:correlations_magnetism}(e) and
\ref{fig:correlations_magnetism}(f) compare configurations in which the
moments remain ferromagnetically aligned within each trimer, while the
relative alignment between neighboring trimers is either ferromagnetic (FM)
or antiferromagnetic (AFM). The FM inter-trimer configuration is
strongly favored, with the AFM arrangement lying
approximately $58$ meV/Cr higher in energy. The two magnetic
configurations also display qualitatively different electronic
structures. The energetically favored FM alignment retains
the characteristic isolated nearly flat band at the Fermi level,
whereas the AFM inter-trimer configuration substantially
reconstructs the bands and no longer exhibits the same isolated
flat-band state. These results show that the flat band is realized in
the magnetically favored FM state, while both intra- and
inter-trimer magnetic interactions strongly favor FM
alignment in the $\sqrt{3}\times\sqrt{3}$ phase.

\subsection*{Electric-field control of the trimerized flat-band phase}

The intrinsic out-of-plane polarity of the
$\sqrt{3}\times\sqrt{3}$ reconstruction provides a natural route to
couple the structural distortion to an external electric field. We
therefore first examine how the electronic structure evolves along a
continuous structural interpolation connecting the high-symmetry 1T
configuration with the fully developed polar
$\sqrt{3}\times\sqrt{3}$ phase.

Figure~\ref{fig:structural_path}(a--d) shows the majority-spin band
structure along this interpolation. At $\lambda=0$, corresponding to
the undistorted 1T structure, the bands around the Fermi level are
strongly dispersive and no isolated narrow band is present. As the
distortion develops, the Cr atoms progressively form the
characteristic trimers, while the Se sublattice undergoes the
accompanying asymmetric out-of-plane reconstruction. At the same time,
the low-energy Cr-$d$ manifold is continuously reorganized and the band
close to the Fermi level progressively loses dispersion, evolving into
the isolated nearly flat band characteristic of the fully developed
$\sqrt{3}\times\sqrt{3}$ phase.

The interpolation therefore shows that the flat-band state develops
together with the full polar CDW reconstruction. As established by the
tight-binding analysis above, the microscopic origin of the narrow band
is the formation of Cr-trimer molecular states. The relevance of the
accompanying polar distortion is that it provides a structural degree
of freedom that can couple to an electric field applied perpendicular
to the monolayer. This motivates us to investigate whether such a field
can provide an external handle on the trimerized flat-band phase.

\begin{figure}[t]
    \centering
    \includegraphics[width=\columnwidth]{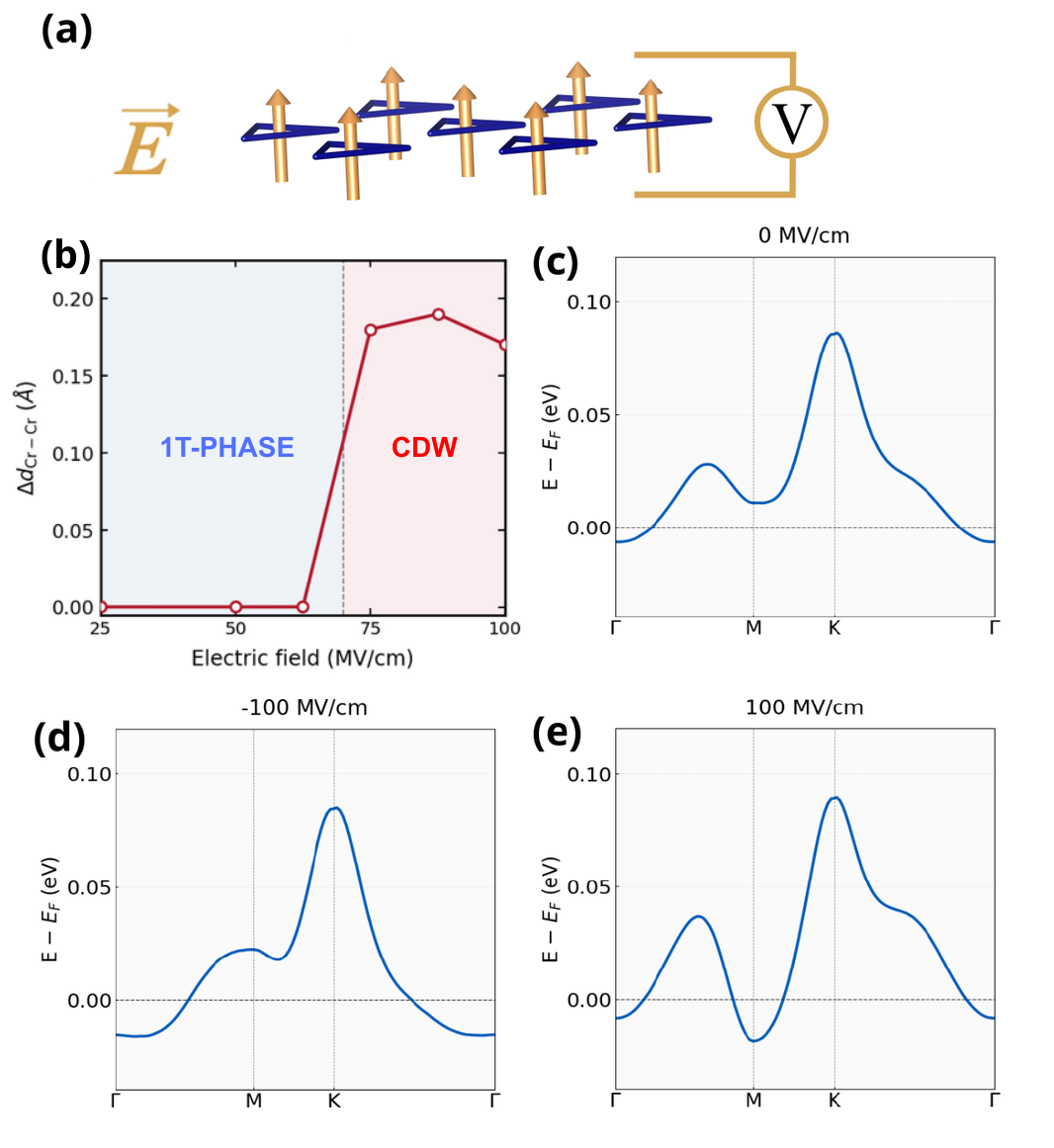}
    \caption{
    Electric-field control of the trimerized flat-band phase.
    (a) Schematic representation of an electric field applied
    perpendicular to the CrSe$_2$ monolayer and along the polar
    structural axis.
    (b) Difference between the long and short Cr--Cr distances,
    $\Delta d_{\mathrm{Cr-Cr}}=d_{\mathrm{long}}-d_{\mathrm{short}}$,
    obtained after structural relaxation starting from the
    high-symmetry 1T configuration at different electric fields.
    For the perfectly symmetric starting geometry used here, a marked
    increase in $\Delta d_{\mathrm{Cr-Cr}}$ appears around
    $70$ MV/cm, signaling the development of the
    $\sqrt{3}\times\sqrt{3}$ trimerized structure. This numerical
    field scale is not interpreted as an intrinsic switching
    threshold.
    (c--e) Evolution of the nearly flat band of the already trimerized
    phase under electric fields of $0$, $-100$, and $+100$ MV/cm,
    respectively, showing the field-induced modification of its
    detailed dispersion.
    }
    \label{fig:electric_field}
\end{figure}

To this end, we performed structural relaxations starting from the
high-symmetry 1T configuration under an electric field applied
perpendicular to the monolayer, as illustrated in
Fig.~\ref{fig:electric_field}(a). Figure~\ref{fig:electric_field}(b)
shows the resulting Cr--Cr bond differentiation,
$\Delta d_{\mathrm{Cr-Cr}}=d_{\mathrm{long}}-d_{\mathrm{short}}$.
For the perfectly symmetric starting geometry used here, the system
remains on the high-symmetry structural branch at lower fields, with
$\Delta d_{\mathrm{Cr-Cr}}$ close to zero. A marked increase appears at fields of approximately
$70$ MV/cm, indicating the development of the Cr trimerization during
the structural relaxation.

Since the parent 1T configuration is dynamically unstable, this field
scale should not be interpreted as an intrinsic switching threshold.
Its value depends on the symmetric initial condition and on the details
of the structural optimization. Rather, these calculations provide a
proof of principle that a perpendicular electric field couples to the
collective polar reconstruction and can promote access to the
$\sqrt{3}\times\sqrt{3}$ trimerized state.

Importantly, once the $\sqrt{3}\times\sqrt{3}$ reconstruction has
formed, reversing the electric field does not destabilize the Cr
trimerization within the range considered here. In particular, a field
opposite to the polar axis does not drive the relaxed structure back
toward the 1T configuration, even though it tends to reduce the polar
distortion. The trimerized CDW therefore remains structurally robust
for both field orientations over the investigated range.

This robustness allows the electric field to be used as a tuning
parameter for the electronic structure once the flat-band phase has
already formed. Figures~\ref{fig:electric_field}(c--e) compare the band
dispersion around the Fermi level at zero field and for fields of
opposite sign. The isolated narrow band remains present throughout this
range, while its detailed dispersion is modified, particularly around
the $M$ and $K$ regions of the Brillouin zone. The electric field can
therefore tune the electronic states of the flat-band phase without
destroying the underlying trimerized reconstruction.

The polar $\sqrt{3}\times\sqrt{3}$ CDW thus provides two complementary
levels of electric-field control. Cr trimerization remains the
microscopic origin of the narrow band, while the accompanying polarity
provides an external handle on the structural reconstruction. Once the
trimerized state is formed, the same field can further modify the
dispersion of the resulting flat band.


In this work, we have investigated the metastable polar
$\sqrt{3}\times\sqrt{3}$ trimerized phase of monolayer CrSe$_2$,
which hosts an isolated, nearly dispersionless band pinned at the
Fermi level in the majority-spin channel. Our results show that this
flat band is a direct consequence of the structural reconstruction:
the formation of Cr trimers reorganizes the low-energy Cr-$d$
manifold into molecular states, with the narrow band arising
predominantly from the in-plane $d_{xy}$ and $d_{x^2-y^2}$ orbitals.
The strong difference between intra- and inter-trimer hybridization
then suppresses the dispersion of this state, providing a simple
microscopic description for the origin of the flat band.

This description also clarifies the roles of electronic correlations
and magnetism. The flat band is already present at the semilocal-DFT
level and therefore does not originate from on-site Coulomb
interactions; increasing the interaction strength instead makes the
trimer-derived narrow band more dispersive. Ferromagnetic order is
energetically favored both within and between Cr trimers, while
preserving the characteristic flat-band state.

The structural reconstruction also provides an intrinsic out-of-plane
polarization. While Cr trimerization is the microscopic origin of the
flat band, the accompanying asymmetric displacement of the Se
sublattice breaks inversion symmetry and enables coupling to a
perpendicular electric field. Our finite-field calculations show that
such a field can promote access to the trimerized phase and tune the
dispersion of the narrow band once formed, while the CDW remains
robust upon field reversal within the investigated range.

The resulting picture is therefore different from the usual routes to
flat-band formation in moir\'e systems or kagome lattices. Here, the
flat band emerges intrinsically from a CDW-driven molecular
reorganization of the Cr $d$ states, while the same structural
reconstruction simultaneously introduces magnetism and an electrically
active polar degree of freedom. Monolayer CrSe$_2$ thus constitutes an
example of how lattice reconstruction can be used not only to generate
flat-band physics, but also to make it externally tunable without the
need for artificial patterning or engineered superlattices.

\section*{Methods}\label{sec:methods}

\subsection*{Electronic and magnetic calculations}

Electronic-structure calculations were performed within the framework
of density functional theory (DFT)
\cite{hohenberg1964inhomogeneous,kohn1965self} using the all-electron
full-potential linearized augmented plane-wave (FP-LAPW) method
as implemented in the \textsc{WIEN2k} code
\cite{blaha2020wien2k}. The exchange--correlation energy was treated
within the generalized gradient approximation (GGA), employing
the Perdew--Burke--Ernzerhof (PBE) functional
\cite{perdew1996generalized}.

For the FP-LAPW calculations, the parameter
$R_{MT}K_{\mathrm{max}}$ was set to \textit{7.0}, and muffin-tin
radii of \textit{2.37} and \textit{2.10} a.u. were used for Cr
and Se atoms, respectively. Brillouin-zone integrations were
performed using a \textit{12$\times$12$\times$1} $k$-point mesh.

On-site Coulomb interactions were introduced using the rotationally
invariant LDA+$U$ approach \cite{petukhov2003correlated}, with
Hubbard $U-J$ values varying from 0 to 5~eV, with $J=0.5$ eV,
in order to assess correlation effects on the electronic and
magnetic properties. Both ferromagnetic and antiferromagnetic
configurations were examined, including non-collinear arrangements
within the $\sqrt{3}\times\sqrt{3}$ supercell, and their relative
energies were monitored as a function of $U$ to determine the
magnetic ground state.

\subsection*{Structural relaxation}

Structural relaxations were performed at the GGA level using
the \textsc{VASP} package
\cite{kresse1996efficient,kresse1996efficiency}, allowing full
optimization of internal coordinates and the in-plane lattice
parameter. The plane-wave energy cutoffs ENCUT and ENAUG were
set to \textit{300 eV} and \textit{550 eV}, respectively.

To model the isolated monolayer, a vacuum spacing of 15~\AA\
was applied along the out-of-plane direction, eliminating
spurious interlayer interactions.

\subsection*{Band structures and unfolding}

Electronic band structures and densities of states were obtained
from the converged charge densities of the \textsc{WIEN2k}
calculations. Unfolded band structures, mapped onto the primitive
$1\times1$ Brillouin zone, were generated to enable a direct
comparison between the distorted CDW phases and the parent
1T structure using
\textsc{Fold2Bloch}\cite{rubel2014unfolding,wang1998majority}.

\subsection*{Electric-field calculations}

Electric-field calculations were performed by applying a
sawtooth-like electrostatic potential along the out-of-plane direction,
while retaining periodic boundary conditions. The monolayer was placed
away from the discontinuity of the potential, within the region where
the potential varies linearly, thereby simulating an approximately
uniform electric field perpendicular to the slab. Structural
relaxations and electronic-structure calculations were then carried out
for different field strengths and orientations.

\subsection*{Tight-binding model}

The effective tight-binding model was implemented using the PythTB
package.\cite{Cole_PythTB_2025} The model contains the Cr
$d_{z^2}$, $d_{xy}$, and $d_{x^2-y^2}$ orbitals on the three Cr sites
of the $\sqrt{3}\times\sqrt{3}$ unit cell and describes the
majority-spin sector of the ferromagnetic calculation. Trimerization
was modeled by distinguishing intra- and inter-trimer hopping
amplitudes, as described in the subsection \hyperref[sec:flat_band_origin]{``Trimerization-driven origin of the flat band''}.
Further details of the three- and five-orbital models and the spin--orbit-coupling analysis are provided in Appendix~A (Figs.~\ref{fig:three_orbital}--\ref{fig:soc}).

\section*{Author contributions}

P.S. and M.C.F.S. carried out most of the calculations. P.S. produced the tight-binding model and all the microscopic description based on it. J.C.S. provided the unfoldings and the calculations of the non-collinear antiferromagnetic phase. J.P. first identified the flat band in this system. J.P., A.O.F., S.B.C. and V.P. conceived the project. V.P. coordinated the project and wrote the first version of the article. P.S., M.C.F.S., J.P., J.C.S. produced and organized the figures. All authors contributed to the discussion and final version of the paper.

\section*{Competing interests}

The authors declare no competing financial or non-financial interests.

\section*{Data Availability}

The datasets generated and/or analyzed during the current study are not publicly available due to the fact that they have not been deposited in a public repository but are available from the corresponding author on reasonable request.

\section*{Acknowledgements}

We thank the CESGA (Centro de Supercomputacion de Galicia) for the computing facilities provided.

\section*{Funding}

V.P. acknowledges support from the Ministry of Science of Spain through the Projects No. PID2021-122609NB-C22 and PID2024-161503NB-C22. S.B.C. acknowledges support from the Ministry of Science of Spain through the Projects No. PID2021-122609NB-C21 and PID2024-161503NB-C21. J.C.S. acknowledges support from the Ministry of Science and Education through the FPU Program (FPU22/01312). A.O.F. acknowledges support from the Academy of Finland Project No. 369367. Xunta de Galicia is acknowledged for projects ED431F 2022/005 and ED431B 2023/055.

The funders provided financial support only and had no role in the study design, data collection and analysis, preparation of the manuscript, or decision to publish.

\FloatBarrier

%

\clearpage
\onecolumngrid
\appendix
\setcounter{section}{1}
\providecommand{\theHfigure}{}
\providecommand{\theHtable}{}
\providecommand{\theHequation}{}
\renewcommand{\theHfigure}{appendix.A.\arabic{figure}}
\renewcommand{\theHtable}{appendix.A.\arabic{table}}
\renewcommand{\theHequation}{appendix.A.\arabic{equation}}
\pdfbookmark[0]{Appendix A: Supplementary Information}{appendix-si}


\thispagestyle{empty}
\begingroup
\renewcommand{\baselinestretch}{1}\selectfont
\vspace*{12pt}
\noindent\begin{minipage}{\textwidth}
\centering
{\sffamily\bfseries\fontsize{11}{14}\selectfont
APPENDIX A: SUPPLEMENTARY INFORMATION\par}
\vspace{22pt}
{\bfseries\fontsize{18}{23}\selectfont
Magnetoelectric flat band induced by a\\[3pt]
$\sqrt{3}\times\sqrt{3}$ charge density wave\\[3pt]
in monolayer CrSe$_2$\par}
\vspace{25pt}
{\fontsize{11}{17}\selectfont
\mbox{Pablo Savino\textsuperscript{1,2}},
\mbox{M. del Carmen Fuente Santiago\textsuperscript{1,2}},\\
\mbox{Jan Phillips\textsuperscript{3}},
\mbox{Javier Corral Sertal\textsuperscript{1,4}},
\mbox{Adolfo O. Fumega\textsuperscript{5}},\\
\mbox{Santiago Blanco-Canosa\textsuperscript{6,7}} and
\mbox{Victor Pardo\textsuperscript{1,2}}\par}
\vspace{19pt}
\rule{\textwidth}{0.4pt}\par
\vspace{15pt}
\begin{minipage}{0.96\textwidth}
\raggedright
\fontsize{10}{13}\selectfont
\begin{list}{}{%
    \setlength{\leftmargin}{13pt}%
    \setlength{\labelwidth}{8pt}%
    \setlength{\labelsep}{5pt}%
    \setlength{\itemsep}{6pt}%
    \setlength{\parsep}{0pt}%
    \setlength{\topsep}{0pt}%
}
\item[\textsuperscript{1}]
Departamento de F\'isica Aplicada, Universidade de Santiago de Compostela,
E-15782 Santiago de Compostela, Spain
\item[\textsuperscript{2}]
Instituto de Materiais iMATUS, Universidade de Santiago de Compostela,
E-15782 Campus Sur s/n, Santiago de Compostela, Spain
\item[\textsuperscript{3}]
International Iberian Nanotechnology Laboratory (INL),
Av. Mestre Jos\'e Veiga, 4715-330 Braga, Portugal
\item[\textsuperscript{4}]
Center for Research in Biological Chemistry and Molecular Materials CIQUS,
Universidade de Santiago de Compostela,
E-15782 Campus Sur s/n, Santiago de Compostela, Spain
\item[\textsuperscript{5}]
Department of Applied Physics, Aalto University, 02150 Espoo, Finland
\item[\textsuperscript{6}]
Donostia International Physics Center (DIPC),
Paseo Manuel de Lardiz\'abal, 20018 San Sebasti\'an, Spain
\item[\textsuperscript{7}]
IKERBASQUE, Basque Foundation for Science, 48013 Bilbao, Spain
\end{list}
\end{minipage}\par
\vspace{23pt}
{\fontsize{10}{14}\selectfont
\textbf{Corresponding author:} Victor Pardo\par
\href{mailto:victor.pardo@usc.es}{victor.pardo@usc.es}\par}
\end{minipage}\par
\endgroup
\clearpage


\renewcommand{\thefigure}{A\arabic{figure}}
\renewcommand{\thetable}{A\arabic{table}}
\renewcommand{\theequation}{A\arabic{equation}}

\setcounter{figure}{0}
\setcounter{table}{0}
\setcounter{equation}{0}


\section*{Supplementary Notes}

\subsection*{Three-orbital effective tight-binding model}

To clarify the microscopic origin of the nearly flat band found in the
majority-spin electronic structure of the
$\sqrt{3}\times\sqrt{3}$ trimerized phase, we construct a minimal
effective tight-binding model using the PythTB package.
\cite{Cole_PythTB_2025}
The purpose of this model is not to provide a quantitatively exact
Wannier interpolation of the first-principles bands, but rather to
identify the minimal orbital and hopping ingredients responsible for
their characteristic organization and, in particular, for the isolated
weakly dispersive band close to the Fermi level.

The $\sqrt{3}\times\sqrt{3}$ unit cell contains three Cr atoms.
Motivated by the orbital-resolved DFT calculations, we initially retain
the $d_{z^2}$, $d_{x^2-y^2}$, and $d_{xy}$ orbitals on each Cr site.
The resulting basis contains nine orbitals,

\begin{equation}
\label{eq:supp_basis_three_orbital}
\left(
d_{z^2}^{1},d_{x^2-y^2}^{1},d_{xy}^{1},
d_{z^2}^{2},d_{x^2-y^2}^{2},d_{xy}^{2},
d_{z^2}^{3},d_{x^2-y^2}^{3},d_{xy}^{3}
\right),
\end{equation}

where the superscript labels the three Cr atoms belonging to a trimer.

The model describes the majority-spin sector of the ferromagnetic
calculation. In the absence of spin--orbit coupling, spin is therefore
not included explicitly and the Hamiltonian should be understood as an
effective single-spin description.

The Hamiltonian is decomposed as

\begin{equation}
\label{eq:supp_hamiltonian_decomposition}
H
=
H_{\mathrm{onsite}}
+
H_{\mathrm{intra}}
+
H_{\mathrm{inter}},
\end{equation}

where the last two terms describe hopping within a Cr trimer and
between neighboring trimers, respectively. The on-site contribution is

\begin{equation}
\label{eq:supp_onsite_hamiltonian}
H_{\mathrm{onsite}}
=
\sum_i
\left[
\varepsilon_{z^2}
c^{\dagger}_{i,z^2}c_{i,z^2}
+
\varepsilon_E
\left(
c^{\dagger}_{i,x^2-y^2}c_{i,x^2-y^2}
+
c^{\dagger}_{i,xy}c_{i,xy}
\right)
\right].
\end{equation}

The two in-plane orbitals are assigned the same effective on-site
energy $\varepsilon_E$, while the $d_{z^2}$ orbital is described by
$\varepsilon_{z^2}$.

For a Cr--Cr bond forming an angle $\phi$ with the reference
$x$ direction, we define

\begin{equation}
\label{eq:supp_angular_factors}
c=\cos(2\phi),
\qquad
s=\sin(2\phi),
\end{equation}

and write the hopping matrix in the basis
$(d_{z^2},d_{x^2-y^2},d_{xy})$ as

\begin{equation}
\label{eq:supp_hopping_three_orbital}
T(\phi)
=
\begin{pmatrix}
t_{z^2}
&
qc
&
qs
\\[4pt]
qc
&
t_{\sigma}c^2+t_{\pi}s^2
&
(t_{\sigma}-t_{\pi})cs
\\[4pt]
qs
&
(t_{\sigma}-t_{\pi})cs
&
t_{\sigma}s^2+t_{\pi}c^2
\end{pmatrix}.
\end{equation}

Here, $t_{\sigma}$ and $t_{\pi}$ describe the two effective hopping
channels within the in-plane
$d_{x^2-y^2}$--$d_{xy}$ manifold, while $t_{z^2}$ describes hopping
between $d_{z^2}$ orbitals. The $2\phi$ angular dependence reflects
the transformation properties of the two real in-plane $d$ orbitals
under rotations within the layer.

The parameter $q$ introduces a bond-dependent hybridization between
the $d_{z^2}$ orbital and the in-plane doublet. For $q=0$, the
$d_{z^2}$ and $d_{x^2-y^2}$--$d_{xy}$ sectors are completely
decoupled. Finite $q$ allows orbital weight to be transferred between
the two sectors and provides a simple phenomenological description of
the orbital mixing present in the full first-principles electronic
structure.

The parameters of the minimal three-orbital model are

\begin{equation}
\label{eq:supp_onsite_three_orbital}
\varepsilon_{z^2}=+0.129~\mathrm{eV},
\qquad
\varepsilon_E=-0.543~\mathrm{eV},
\end{equation}

with intra-trimer hoppings

\begin{equation}
\label{eq:supp_intra_three_orbital}
\begin{aligned}
t_{z^2}^{\mathrm{intra}}&=-0.343~\mathrm{eV},
\\
t_{\sigma}^{\mathrm{intra}}&=-0.370~\mathrm{eV},
\\
t_{\pi}^{\mathrm{intra}}&=+0.056~\mathrm{eV},
\end{aligned}
\end{equation}

and inter-trimer hoppings

\begin{equation}
\label{eq:supp_inter_three_orbital}
\begin{aligned}
t_{z^2}^{\mathrm{inter}}&=-0.113~\mathrm{eV},
\\
t_{\sigma}^{\mathrm{inter}}&=-0.001~\mathrm{eV},
\\
t_{\pi}^{\mathrm{inter}}&=+0.036~\mathrm{eV}.
\end{aligned}
\end{equation}

The strong difference between intra- and inter-trimer hopping reflects
the bond differentiation generated by the structural trimerization.

The molecular origin of the band organization can be seen most clearly
in the isolated-trimer limit. Considering first the $d_{z^2}$ sector,
the Hamiltonian of one triangular Cr trimer is

\begin{equation}
\label{eq:supp_isolated_trimer_hamiltonian}
H_{z^2}^{\mathrm{trimer}}
=
\begin{pmatrix}
\varepsilon_{z^2} & t_{z^2} & t_{z^2} \\
t_{z^2} & \varepsilon_{z^2} & t_{z^2} \\
t_{z^2} & t_{z^2} & \varepsilon_{z^2}
\end{pmatrix}.
\end{equation}

Its spectrum contains one fully symmetric molecular state at

\begin{equation}
\label{eq:supp_trimer_singlet}
E_{\mathrm{s}}
=
\varepsilon_{z^2}
+
2t_{z^2},
\end{equation}

and a doubly degenerate pair at

\begin{equation}
\label{eq:supp_trimer_doublet}
E_{\mathrm{d}}
=
\varepsilon_{z^2}
-
t_{z^2}.
\end{equation}

For the parameters used here, these levels occur at approximately
$-0.557$ and $+0.472$ eV, respectively.

The in-plane sector contains six molecular states arising from two
orbitals on each of the three Cr sites. The directional
$\sigma$- and $\pi$-like hopping channels reorganize them into a
characteristic $2+1+2+1$ molecular-level sequence. For the present
parameters, the corresponding energies in the isolated-trimer limit
are approximately $-0.939$, $-0.812$, $-0.304$, and $+0.040$ eV.

The upper in-plane molecular state is particularly relevant because it
lies close to the Fermi level. When the coupling between neighboring
trimers is restored, this molecular state acquires only a relatively
small dispersion and evolves into the isolated narrow band found in
the periodic model. The flat-band mechanism can therefore be
understood as the formation of a molecular state within each Cr
trimer followed by weak propagation between neighboring trimers.

\begin{figure}[t]
    \centering
    \includegraphics[width=\textwidth]{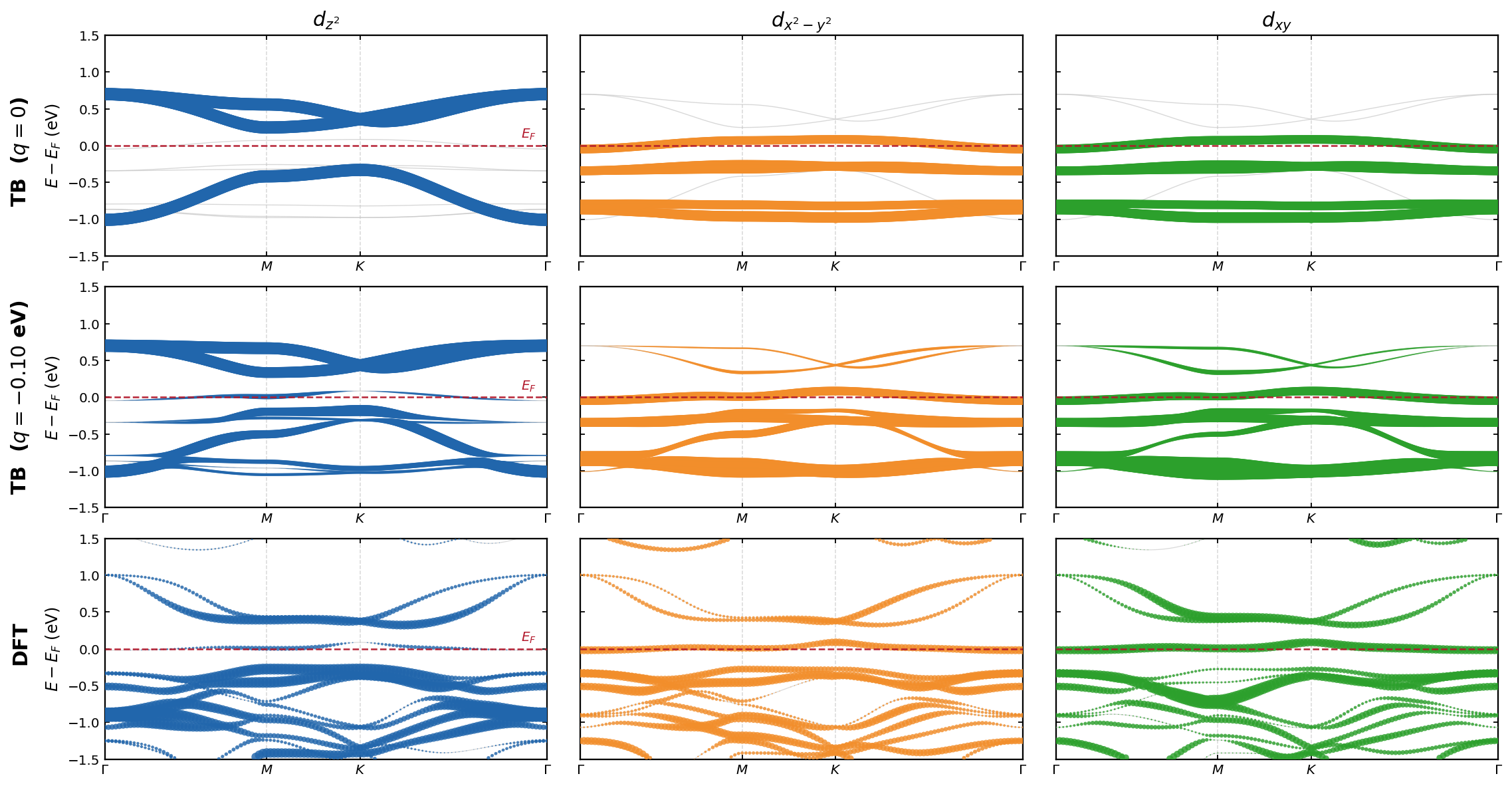}
    \caption{
    Orbital-resolved comparison between the three-orbital effective
    tight-binding model and the majority-spin DFT electronic structure
    of the $\sqrt{3}\times\sqrt{3}$ trimerized phase.
    The columns show the $d_{z^2}$, $d_{x^2-y^2}$, and $d_{xy}$
    orbital weights.
    The upper row corresponds to the decoupled limit $q=0$.
    The middle row shows the model for $q=-0.10$ eV, where finite
    $d_{z^2}$--in-plane hybridization produces a redistribution of
    orbital weight.
    The lower row shows the corresponding majority-spin DFT fat bands.
    The isolated weakly dispersive band close to the Fermi level is
    already present for $q=0$ and remains predominantly derived from
    the in-plane $d_{x^2-y^2}$ and $d_{xy}$ orbitals after finite
    hybridization is introduced.
    }
    \label{fig:three_orbital}
\end{figure}

Figure~\ref{fig:three_orbital} illustrates the orbital structure of
this model and the role of $q$. The upper row corresponds to the
decoupled limit $q=0$. In this case, the $d_{z^2}$ and in-plane
orbital manifolds are completely separated. Importantly, the isolated
weakly dispersive band around the Fermi level is already present and
carries predominantly $d_{x^2-y^2}$ and $d_{xy}$ character. The
existence of this band therefore does not require
$d_{z^2}$--in-plane hybridization.

The middle row shows the same model for $q=-0.10$ eV. A finite $q$
redistributes orbital weight between the $d_{z^2}$ and in-plane
manifolds and substantially increases their hybridization. Several
bands that are pure $d_{z^2}$ or pure in-plane states in the
$q=0$ limit acquire mixed character. Nevertheless, the isolated band
close to the Fermi level remains predominantly derived from the
$d_{x^2-y^2}$ and $d_{xy}$ orbitals.

The lower row shows the corresponding majority-spin DFT orbital
projections. Although the first-principles electronic structure
contains additional Cr-$d$ and Se-$p$ states absent from the minimal
model, the same qualitative orbital hierarchy is evident. In
particular, the isolated state at the Fermi level is dominated by the
in-plane $d_{x^2-y^2}$ and $d_{xy}$ orbitals, whereas the
$d_{z^2}$ states contribute more strongly to the surrounding
dispersive bands.

The comparison therefore separates two distinct effects. Trimerization
and the resulting molecular organization of the in-plane orbitals
generate the isolated narrow band, while the parameter $q$ describes
additional mixing between orbital manifolds. The latter modifies the
detailed orbital distribution but is not responsible for the
appearance of the flat band itself.

The minimal mechanism may therefore be summarized as

\begin{equation}
\label{eq:supp_flat_band_mechanism}
\begin{gathered}
\text{Cr trimerization}
\longrightarrow
\text{in-plane trimer molecular state}
\\
\longrightarrow
\text{weak inter-trimer coupling}
\longrightarrow
\text{isolated narrow band}.
\end{gathered}
\end{equation}


\subsection*{Extension to the full Cr-$3d$ manifold}

The three-orbital model is sufficient to identify the microscopic
mechanism responsible for the isolated narrow band. Nevertheless, the
first-principles electronic structure contains contributions from the
complete Cr-$3d$ manifold. To test whether the mechanism survives when
these additional degrees of freedom are included, we extend the model
by explicitly adding the $d_{xz}$ and $d_{yz}$ orbitals on every Cr
site.

The orbital basis of the extended model is therefore

\begin{equation}
\label{eq:supp_basis_five_orbital}
\left(
d_{z^2},
d_{x^2-y^2},
d_{xy},
d_{xz},
d_{yz}
\right)
\end{equation}

on each of the three Cr atoms in the
$\sqrt{3}\times\sqrt{3}$ cell, giving a total of fifteen orbitals.

The original $d_{z^2}$ and
$d_{x^2-y^2}$--$d_{xy}$ hopping structure is retained. The additional
$d_{xz}$--$d_{yz}$ sector forms a second directional doublet. Whereas
the in-plane $d_{x^2-y^2}$--$d_{xy}$ pair transforms with $2\phi$,
the $(d_{xz},d_{yz})$ pair transforms with the bond angle $\phi$.
Its hopping matrix is written as

\begin{equation}
\label{eq:supp_hopping_xz_yz}
T_{xz/yz}(\phi)
=
\begin{pmatrix}
t_{\sigma}^{xz/yz}\cos^2\phi
+
t_{\pi}^{xz/yz}\sin^2\phi
&
\left(
t_{\sigma}^{xz/yz}
-
t_{\pi}^{xz/yz}
\right)
\cos\phi\sin\phi
\\[4pt]
\left(
t_{\sigma}^{xz/yz}
-
t_{\pi}^{xz/yz}
\right)
\cos\phi\sin\phi
&
t_{\sigma}^{xz/yz}\sin^2\phi
+
t_{\pi}^{xz/yz}\cos^2\phi
\end{pmatrix}.
\end{equation}

As in the minimal model, the intra- and inter-trimer hopping amplitudes
are allowed to differ. The $d_{xz}$ and $d_{yz}$ on-site energies are
placed below the Fermi-level region, consistently with the
first-principles orbital projections, where these states contribute
predominantly to the lower-energy Cr-$d$ manifold.

For the extended model we use

\begin{equation}
\label{eq:supp_onsite_xz_yz}
\varepsilon_{xz/yz}=-0.820~\mathrm{eV},
\end{equation}

with intra-trimer hoppings

\begin{equation}
\label{eq:supp_intra_xz_yz}
t_{\sigma,xz/yz}^{\mathrm{intra}}
=
-0.240~\mathrm{eV},
\qquad
t_{\pi,xz/yz}^{\mathrm{intra}}
=
+0.090~\mathrm{eV},
\end{equation}

and inter-trimer hoppings

\begin{equation}
\label{eq:supp_inter_xz_yz}
t_{\sigma,xz/yz}^{\mathrm{inter}}
=
-0.045~\mathrm{eV},
\qquad
t_{\pi,xz/yz}^{\mathrm{inter}}
=
+0.025~\mathrm{eV}.
\end{equation}

A small effective mixing between $d_{z^2}$ and the
$d_{xz}$--$d_{yz}$ sector is also retained. Together with the finite
$d_{z^2}$--in-plane mixing already introduced in the three-orbital
description, this allows the model to reproduce the more complex
orbital redistribution present in the DFT calculation.

\begin{figure}[t]
    \centering
    \includegraphics[width=\textwidth]{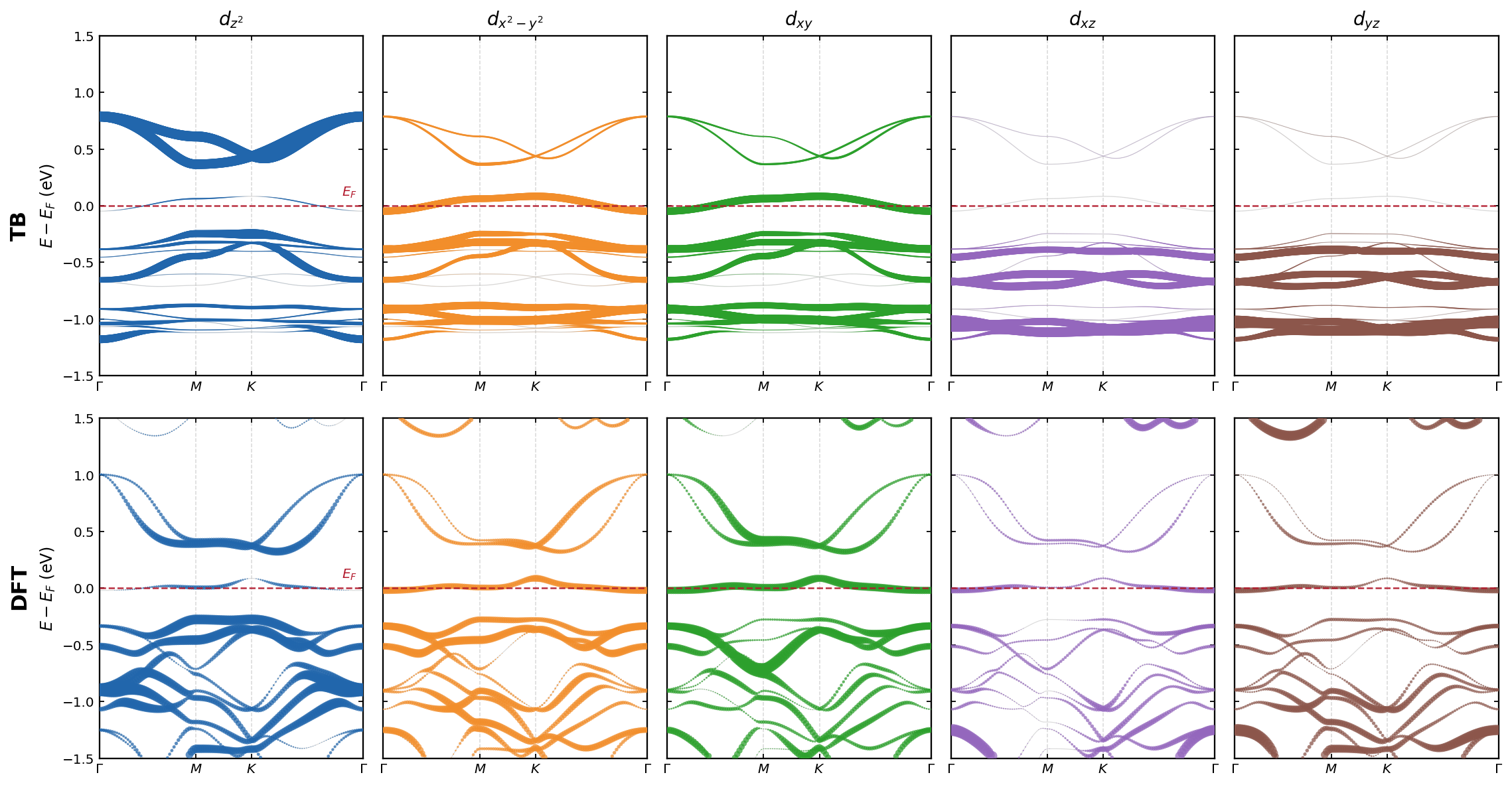}
    \caption{
    Orbital-resolved comparison between the extended five-orbital
    tight-binding model and the majority-spin DFT electronic structure
    of the $\sqrt{3}\times\sqrt{3}$ trimerized phase.
    The columns correspond to $d_{z^2}$, $d_{x^2-y^2}$, $d_{xy}$,
    $d_{xz}$, and $d_{yz}$, respectively.
    The upper row shows the effective-model orbital weights and the
    lower row shows the corresponding DFT fat bands.
    The full five-orbital model reproduces the main orbital hierarchy
    obtained from first principles: the isolated narrow state at the
    Fermi level is predominantly composed of the in-plane
    $d_{x^2-y^2}$ and $d_{xy}$ orbitals, whereas the
    $d_{xz}$ and $d_{yz}$ states contribute mainly to the occupied
    lower-energy manifold.
    }
    \label{fig:five_orbital}
\end{figure}

Figure~\ref{fig:five_orbital} compares the orbital-resolved
five-orbital model directly with the majority-spin DFT bands.
The upper row shows the tight-binding orbital weights, while the lower
row shows the corresponding DFT fat bands for all five Cr-$d$
orbitals.

A notable improvement with respect to the minimal model is obtained in
the electronic manifold below the Fermi level. The addition of the
$d_{xz}$ and $d_{yz}$ orbitals generates new molecular bands in the
same energy range where these orbitals carry significant weight in the
first-principles calculation. The extended model therefore captures
not only the low-energy flat-band sector but also much more of the
overall Cr-$d$ band organization.

Most importantly, the orbital character of the states surrounding the
Fermi level remains essentially unchanged. In both the effective model
and the DFT calculation, the isolated narrow band is predominantly
derived from $d_{x^2-y^2}$ and $d_{xy}$. The $d_{xz}$ and $d_{yz}$
states contribute primarily at lower energies, whereas the
$d_{z^2}$ orbital has appreciable weight in the more dispersive bands
above the flat-band region and in part of the occupied manifold.

The close correspondence between the orbital weights of the extended
model and the DFT bands provides an important consistency check of the
molecular-orbital interpretation. In particular, the flat band is not
an artifact of restricting the basis to three orbitals. It remains
present after the complete Cr-$3d$ manifold is included.

The five-orbital calculation also clarifies the different roles of the
orbital sectors. The $d_{x^2-y^2}$--$d_{xy}$ manifold contains the
essential trimer molecular state that evolves into the isolated narrow
band. The $d_{z^2}$ states hybridize with this manifold and contribute
to the surrounding dispersive electronic structure, while the
$d_{xz}$ and $d_{yz}$ states complete the lower-energy Cr-$d$
manifold.

The minimal three-orbital model therefore isolates the microscopic
mechanism, while the extended five-orbital calculation demonstrates
that this mechanism remains valid in a substantially more complete
description of the Cr-$3d$ electronic structure.

The tight-binding parameters in both models were chosen
phenomenologically to reproduce the qualitative orbital hierarchy,
molecular-level structure, and characteristic band dispersions of the
DFT calculation. They were not obtained from a direct Wannier
projection or from a numerical fit of the complete first-principles
Hamiltonian.


\subsection*{Spin--orbit coupling in the effective model}

The fully relativistic DFT calculation shows that spin--orbit coupling
does not qualitatively modify the molecular band structure of the
trimerized phase, but it does lift selected band degeneracies. The
splittings are particularly visible in regions where two nearby bands
carry substantial $d_{x^2-y^2}$ and $d_{xy}$ character. This behavior
can be understood naturally within the same three-orbital effective
model.

The atomic spin--orbit interaction is

\begin{equation}
\label{eq:supp_soc_atomic}
H_{\mathrm{SOC}}
=
\lambda_{\mathrm{SOC}}
\mathbf{L}\cdot\mathbf{S}.
\end{equation}

The effective model describes the majority-spin sector of a
ferromagnetic state. For a spin quantization axis along $z$, and
neglecting spin-flip processes between majority- and minority-spin
states, the relevant projected contribution is

\begin{equation}
\label{eq:supp_soc_projected}
H_{\mathrm{SOC}}^{\mathrm{eff}}
=
\lambda_{\mathrm{SOC}} L_zS_z.
\end{equation}

Taking $S_z=+1/2$ gives

\begin{equation}
\label{eq:supp_soc_spin_up}
H_{\mathrm{SOC}}^{\mathrm{eff}}
=
\frac{\lambda_{\mathrm{SOC}}}{2}L_z.
\end{equation}

The importance of this term becomes clear by considering the action of
$L_z$ on the real $d$-orbital basis. The orbital $d_{z^2}$ has
$m=0$ and therefore satisfies

\begin{equation}
\label{eq:supp_lz_dz2}
L_z|d_{z^2}\rangle=0.
\end{equation}

By contrast, $d_{x^2-y^2}$ and $d_{xy}$ are not individually
eigenstates of $L_z$. Their complex combinations

\begin{equation}
\label{eq:supp_d_plus_two}
|d_{+2}\rangle
=
\frac{1}{\sqrt{2}}
\left(
|d_{x^2-y^2}\rangle
+
i|d_{xy}\rangle
\right),
\end{equation}

\begin{equation}
\label{eq:supp_d_minus_two}
|d_{-2}\rangle
=
\frac{1}{\sqrt{2}}
\left(
|d_{x^2-y^2}\rangle
-
i|d_{xy}\rangle
\right),
\end{equation}

carry opposite orbital angular momentum,
$m=\pm2$, up to the phase convention adopted for the real-orbital
basis.

In the restricted basis

\begin{equation}
\label{eq:supp_basis_soc}
\left(
d_{z^2},
d_{x^2-y^2},
d_{xy}
\right),
\end{equation}

the $L_z$ operator is represented by

\begin{equation}
\label{eq:supp_lz_matrix}
L_z
=
\begin{pmatrix}
0 & 0 & 0 \\
0 & 0 & -2i \\
0 & 2i & 0
\end{pmatrix}.
\end{equation}

The corresponding projected SOC term on each Cr site is therefore

\begin{equation}
\label{eq:supp_soc_matrix}
H_{\mathrm{SOC}}^{\mathrm{eff}}
=
\frac{\lambda_{\mathrm{SOC}}}{2}
\begin{pmatrix}
0 & 0 & 0 \\
0 & 0 & -2i \\
0 & 2i & 0
\end{pmatrix}.
\end{equation}

The imaginary off-diagonal terms are essential. In the absence of SOC,
the effective hopping Hamiltonian can be written entirely in terms of
real matrix elements in the chosen real-orbital basis. SOC couples
$d_{x^2-y^2}$ and $d_{xy}$ through imaginary matrix elements because
the eigenstates of orbital angular momentum are complex combinations
of these two real orbitals.

The appearance of imaginary matrix elements does not imply complex
energy eigenvalues. The SOC Hamiltonian remains Hermitian because

\begin{equation}
\label{eq:supp_soc_hermiticity}
\left(H_{\mathrm{SOC}}\right)_{xy,x^2-y^2}
=
\left[
\left(H_{\mathrm{SOC}}\right)_{x^2-y^2,xy}
\right]^*,
\end{equation}

and its eigenvalues are therefore strictly real.

For an ideal degenerate
$d_{x^2-y^2}$--$d_{xy}$ pair, diagonalizing this SOC block produces
the $m=\pm2$ combinations and shifts them in opposite directions. In
the present convention the corresponding shifts are

\begin{equation}
\label{eq:supp_soc_energy_shifts}
\Delta E_{\pm}
=
\pm \lambda_{\mathrm{SOC}},
\end{equation}

giving an ideal splitting

\begin{equation}
\label{eq:supp_soc_splitting}
\Delta_{\mathrm{SOC}}
=
2\lambda_{\mathrm{SOC}}.
\end{equation}

The periodic trimerized system is more complex because its bands are
molecular states rather than pure atomic orbitals. The SOC-induced
splitting consequently depends on both their in-plane orbital weight
and their pre-existing energy separation. Nevertheless, the basic
selection rule remains visible: regions containing two nearby states
with strong $d_{x^2-y^2}+d_{xy}$ character are the regions most
strongly affected by SOC.

\begin{figure}[t]
    \centering
    \includegraphics[width=\textwidth]{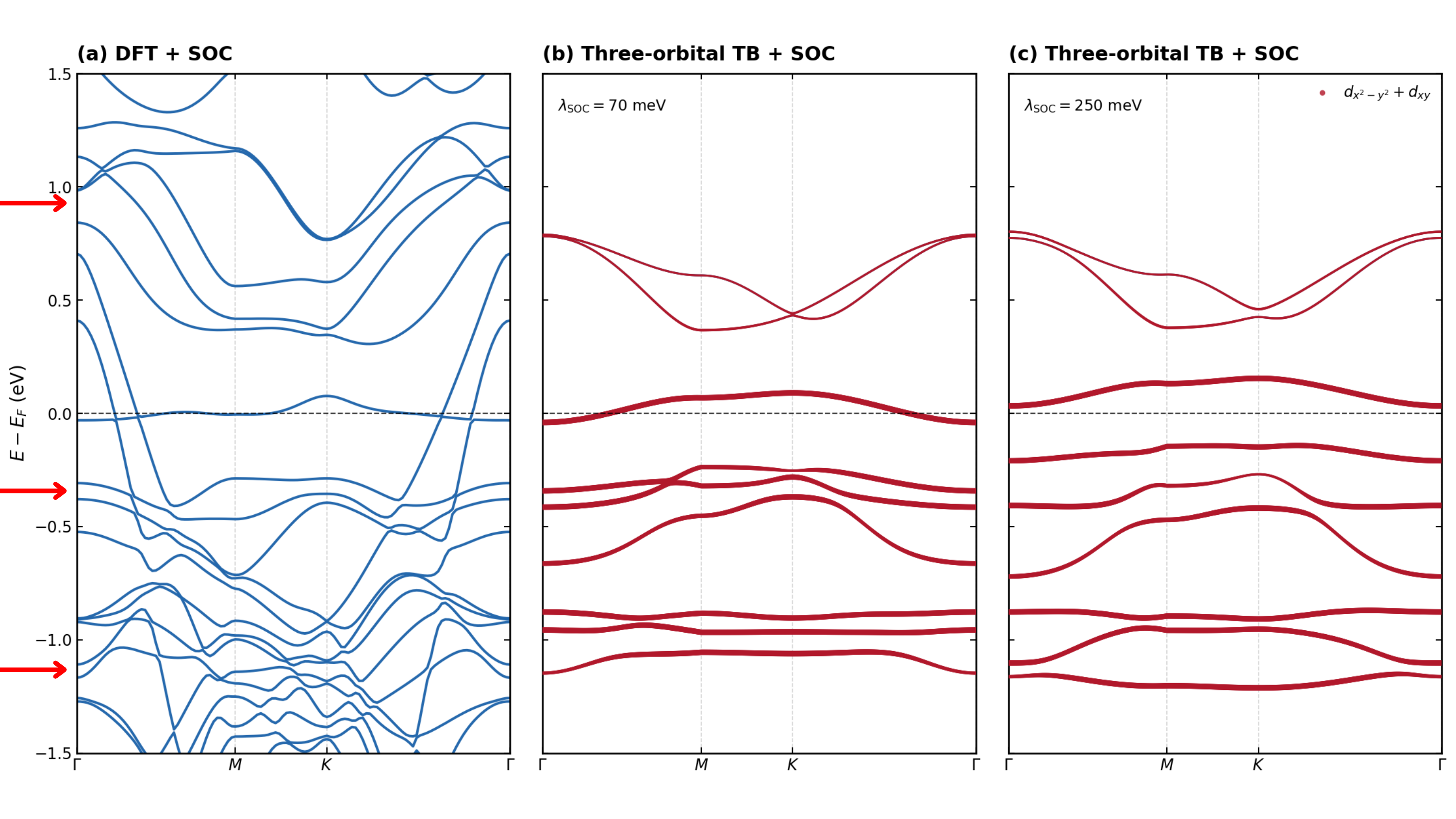}
    \caption{
    Effect of spin--orbit coupling on the trimerized electronic
    structure.
    (a) Fully relativistic DFT band structure of the
    $\sqrt{3}\times\sqrt{3}$ phase.
    (b) Three-orbital tight-binding model including the projected
    SOC interaction with $\lambda_{\mathrm{SOC}}=70$ meV.
    (c) Same model using the deliberately enhanced value
    $\lambda_{\mathrm{SOC}}=250$ meV.
    In the effective-model panels, the red weight denotes the combined
    $d_{x^2-y^2}+d_{xy}$ orbital contribution. The largest
    SOC-induced splittings occur between bands carrying substantial
    in-plane orbital character, reflecting the direct coupling of
    $d_{x^2-y^2}$ and $d_{xy}$ through the $L_z$ operator.
    The $250$ meV calculation is shown only to emphasize this
    orbital-selective mechanism and is not intended as a fitted
    microscopic SOC parameter for CrSe$_2$.
    }
    \label{fig:soc}
\end{figure}

This behavior is shown in Fig.~\ref{fig:soc}. Panel (a) presents the
fully relativistic DFT bands. Several pairs of bands exhibit visible
SOC-induced splittings, while the overall electronic structure and the
narrow state at the Fermi level remain intact.

Panels (b) and (c) show the corresponding three-orbital tight-binding
model after introducing the projected SOC term. The red weight
indicates the total in-plane orbital contribution,

\begin{equation}
\label{eq:supp_inplane_weight}
w_{\parallel}
=
w_{x^2-y^2}
+
w_{xy}.
\end{equation}

The regions where two bands with strong in-plane character approach
one another are precisely those in which the SOC term produces the
largest separations.

For $\lambda_{\mathrm{SOC}}=70$ meV, shown in
Fig.~\ref{fig:soc}(b), the in-plane molecular bands develop clear
splittings while the overall band hierarchy remains essentially
unchanged. In particular, the isolated narrow band survives.

To make the orbital origin of the splitting especially transparent,
Fig.~\ref{fig:soc}(c) shows the same calculation using the deliberately
enhanced value $\lambda_{\mathrm{SOC}}=250$ meV. This value is not
intended as a quantitative estimate of the SOC strength of CrSe$_2$.
It is used only as a numerical demonstration of the structure of the
SOC term. Increasing $\lambda_{\mathrm{SOC}}$ strongly amplifies the
separation between the in-plane-derived molecular states, whereas
states with predominantly $d_{z^2}$ character remain much less
affected at first order.

The SOC analysis therefore provides an additional consistency check of
the orbital interpretation of the narrow band. Trimerization first
produces the molecular electronic structure and the isolated
in-plane-derived flat band. Spin--orbit coupling subsequently acts on
this already formed molecular manifold, mixing the
$d_{x^2-y^2}$ and $d_{xy}$ components and lifting degeneracies where
the corresponding states possess finite orbital angular momentum.

Within this picture, SOC controls the fine structure of the
in-plane-derived bands but is not responsible for the appearance of
the flat band itself. The weakly dispersive state remains fundamentally
a consequence of the molecular reconstruction induced by Cr
trimerization and the weak coupling between neighboring trimers.

\end{document}